\documentclass{elsart}

\usepackage{graphicx}

\usepackage{amssymb}

\begin{document}

\begin{frontmatter}



\title{Coexistence of single--mode and multi--longitudinal mode emission in the ring
laser model}


\author[terrassa]{J.L. Font},
\ead{Josep.Lluis.Font@upc.es}
\author[terrassa]{R. Vilaseca},
\ead{Ramon.Vilaseca@upc.es}
\author[como]{F. Prati\corauthref{cor}},
\corauth[cor]{Corresponding author.}
\ead{franco.prati@uninsubria.it}
\author[valencia]{E. Rold\'{a}n}
\ead{eugenio.roldan@uv.es}

\address[terrassa]{Departament de F\'{\i}sica i Enginyeria Nuclear,
Universitat Polit\`{e}cnica de Catalunya, Colom 11, E--08222
Terrassa, Spain}
\address[como]{INFM--CNR and Dipartimento di Fisica e Matematica,
Universit\`{a} dell'Insubria, via Valleggio 11, I--22100 Como,
Italy}
\address[valencia]{Departament d'\`{O}ptica, Universitat de Val\`{e}ncia, Dr. Moliner 50,
E--46100 Burjassot, Spain}

\begin{abstract}
A homogeneously broadened unidirectonal ring laser can emit in
several longitudinal modes for large enough pump and cavity length
because of Rabi splitting induced gain. This is the so called
Risken-Nummedal--Graham-Haken (RNGH) instability. We investigate
numerically the properties of the multi--mode solution. We show
that this solution can coexist with the single--mode one, and its
stability domain can extend to pump values smaller than the
critical pump of the RNGH instability. Morevoer, we show that the
multi--mode solution for large pump values is affected by two
different instabilities: a pitchfork bifurcation, which preserves
phase--locking, and a Hopf bifurcation, which destroys it.
\end{abstract}

\end{frontmatter}

The Risken-Nummedal--Graham-Haken (RNGH) instability was first
described in two independent papers in 1968 \cite{RN68a,GH68}. In
short, the instability consists in the destabilization of single
longitudinal--mode emission, which appears immediately above the
lasing threshold in a single transverse--mode
homogeneously--broadened unidirectional ring laser, in favor of
multilongitudinal mode emission. The physical mechanism
responsible for that instability is the Rabi splitting of the
lasing transition, induced by the lasing mode, which leads to the
appearance of gain for the sideband modes
\cite{Ikeda89,WeissVilaseca,Khanin}.

After the instability, the laser emits in a pulsing regime because
of the beating between different longitudinal modes, which are
phase--locked. It is to be emphasized that no inhomogeneity, nor
spectral nor spatial, is needed for multi--mode emission. For a
recent review of the RNGH instability, see \cite{Roldan05} and
references therein.

Although some analytical work can be done regarding what happens
after the instability occurs, it is evident that the multi--mode
emission regime needs to be analyzed numerically. The situation is
similar to that of the Lorenz--Haken (LH) instability, which is
the single--mode counterpart of the RNGH instability. However,
while the dynamics associated with the LH instability has been
completely and since a long time characterized through a large
number of numerical studies \cite{Sparrow,Sadiki}, a lot of work
has still to be done to achieve the same degree of knowledge for
the RNGH instability.

The first numerical study about the dynamics associated with the
RNGH instability was carried out by Risken and Nummedal themselves
\cite{RN68b} but since then, along almost forty years, only a few
works have been devoted to that
\cite{Ikeda89,Mayr81,Zorell81,Lugiato85a,Casini97,deValcarcel03a}.
It must also be noted that some of these studies are quite
superficial as they were intended to show some examples of the
pulsing regime rather than characterizing it.

An important aspect of the RNGH instability is the supercritical
or subcritical character of the bifurcation. We remind that if the
pump parameter $A$ is the control parameter and $A_{i}$ is the
pump value at which the single--mode solution destabilizes (also
known as the second laser threshold), the bifurcation is
supercritical if the multi--mode solution that arises from the
instability exists only for $A>A_{i}$, and it is subcritical if it
exists also for $A<A_{i}$. In the latter case, the multi--mode
solution will coexist with the stable single--mode solution in the
interval $A_{sub}\leq A\leq A_{i}$, where $A_{sub}$ must be in
general determined numerically. This region of the parameter space
is called hard excitation domain, because within that domain a
large perturbation of the stable single--mode solution allows to
make the transition to the multi--mode solution.

The understanding of this question may be important for the
correct interpretation of the experimental results recently
obtained in erbium--doped
fiber lasers (EDFLs) \cite{F95,P97,Voigt01,Voigt04} (see also \cite{Roldan05}%
). In fact, if the bifurcation is subcritical the self--pulsing
regime may be observed experimentally for pump values smaller than
the instability threshold $A_{i}$ given by the linear stability
analysis of the single--mode solution, and the transition from cw
emission to self--pulsing would be discontinuous.

\begin{figure}[ptb]
\centering
{\scalebox{.4}{\includegraphics*{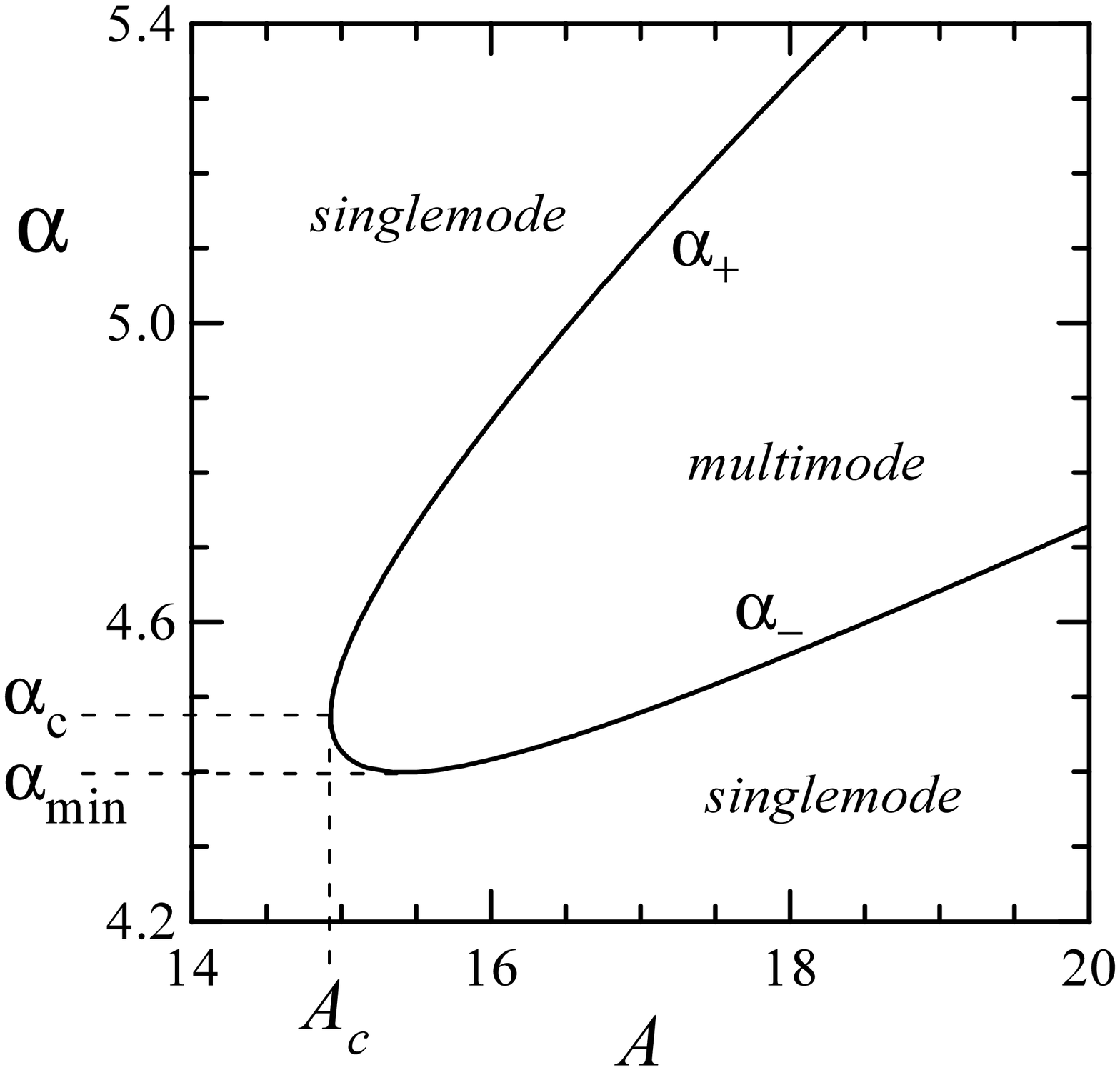}}}\caption{Instability
domain of the single--mode solution for $\gamma=1$ and
$\sigma=0.05$. For this choice of the parameters we have
$A_{c}=14.93$, $\alpha_{c}=4.47$, and
$\alpha_{\mathit{min}}=4.40$.} \label{fig:insta} \vglue1cm
\end{figure}

For the RNGH instability the instability domain is usually
represented in the $\langle A,\alpha\rangle$ plane, where $\alpha$
is the properly scaled side--mode frequency. The instability
domain has the shape of a tongue delimited by the curves
$\alpha_{-}$ and $\alpha_{+}$, which merge at the critical point
$(A_{c},\alpha_{c})$ (see Fig. \ref{fig:insta}). $\alpha_{c}$ is
the critical frequency at which the instability threshold $A_{i}$
attains its minimum value $A_{c}$. The single--mode solution is
unstable if, for a given pump $A$, there is at least one
side--mode whose frequency $\alpha$ lies inside the tongue.

A number of numerical and analytical studies have shown that the
multi--mode solution exists not only inside the instability
domain, but also for $\alpha<\alpha_{-}$ and $A>A_{c}$, where the
linear stability analysis predicts stable single--mode emission.

This result was found numerically already by Risken and Nummedal
in their second paper of 1968 \cite{RN68b}. They showed that,
fixing $A>A_{c}$ and decreasing $\alpha$ from an initial value
larger than $\alpha_{+}$, the single--mode solution bifurcates to
the multi--mode solution at $\alpha_{+}$ and this solution
persists even when the other boundary $\alpha_{-}$ is crossed.
Risken and Nummedal were not able to determine the lower boundary
of the existence domain of the multi--mode solution, because the
numerical analysis of that solution is problematic for small
$\alpha$. In fact, as $\alpha$ decreases the pulses becomes higher
and narrower and in order to reproduce them correctly a very small
spatial stepsize is needed, which implies increasing computation
time.

Later on, Haken and Ohno \cite{Haken76,Ohno76,Haken78} derived a
generalized Ginzburg--Landau equation for the critical (unstable)
mode and found again the coexistence of the two solutions, which
are the minima of an effective potential. They also showed that
the transition from single--mode to multi--mode emission can be
supercritical or subcritical depending on the frequency $\alpha$,
but they did not write an analytic expression of this result.

Simpler analytic results can be found considering some particular
limits for the parameters
$\gamma=\sqrt{\gamma_{\|}/\gamma_{\bot}}$ and $\sigma
=\kappa/\sqrt{\gamma_{\|}\gamma_{\bot}}$, where $\kappa$,
$\gamma_{\bot}$ and $\gamma_{\|}$ are the decay rates of electric
field, medium polarization and population inversion.

In the limit of class--B lasers ($\gamma\ll1,\sigma\approx1$), for
which $A_{c}=9$, Fu \cite{Fu89} derived analytically an
unambiguous condition: If the pump parameter $A$ is the
bifurcation parameter, the bifurcation is supercritical
(subcritical) when $\alpha>\alpha_{c}$ ($\alpha<\alpha_{c}$). This
result has been recently generalized for conditions outside the
uniform field limit \cite{Roldan05}. The same result was found by
Carr and Erneux in a slightly different limit for class--B lasers
($\gamma\ll1,\sigma\gg1$) \cite{Carr94a}.

All these results mean that multi--mode emission can be found for
parameter settings for which the single--mode solution is still
stable. Nevertheless, the minimum instability threshold pump
$A_{c}=9$ has been always found to be a lower bound for
multi--mode emission. In general, the condition that determines
sub-- or supercritical bifurcation is not known, and it remains to
be determined whether the multi--mode solution can exist not only
for $\alpha<\alpha_{-}$, but also for $A<A_{c}$ outside the
class--B limit.

In this paper we address this question and show that, outside the
class B--limit, the multi--mode solution can indeed exist for
$A<A_{c}$ although not below the limit $A=9$. Moreover, we perform
an accurate study of the multi--mode solution and show that,
increasing the pump power $A$ for a fixed frequency $\alpha$, the
solution is affected by two instabilities in sequence. The first
is a pitchfork instability, which breaks the symmetry of the
solution, but preserves phase--locking. The second is a Hopf
instability, which unlock the phases, introducing a slow
modulation of the pulses.

In Section 2 we introduce the model equations and recall the main
results concerning the RNGH instability, in Section 3 we
illustrate and comment the numerical results and finally in
Section 4 we draw our conclusions.

\section{Model}

Consider an incoherently pumped and homogeneously broadened
two--level active medium of length $L_{\mathrm{m}}$, contained in
a ring cavity of length $L_{\mathrm{c}}$, interacting with a
unidirectional plane wave laser field. We assume that the cavity
is resonant with the atomic transition frequency and that the
cavity mirrors reflectivities are close to unity so that the
uniform field limit holds. The Maxwell--Bloch equations describing
such a laser can be written in the form \cite{Roldan05}
\begin{eqnarray}
\left(  \partial_{\tau}+\partial_{\zeta}\right)  F(\zeta,\tau)
&=&\sigma\left(  A\,P-F\right)  \,,\label{mod1}\\
\partial_{\tau}P(\zeta,\tau)  &=& \gamma^{-1}\left(  FD-P\right)
\,,\label{mod2}\\
\partial_{\tau}D(\zeta,\tau)  &=& \gamma\left[  1-D-\operatorname{Re}\left(
FP\right)  \right]  \,. \label{mod3}%
\end{eqnarray}
In these equations $F(\zeta,\tau)$ is the normalized slowly
varying envelope of the laser field and $P(\zeta,\tau)$ and
$D(\zeta,\tau)$ are the normalized slowly varying envelope of the
medium polarization and the population inversion, respectively
(see \cite{Roldan05} for the normalizations). The parameters $A$,
$\sigma$ and $\gamma$ have been already defined in the
Introduction. We use the adimensional time $\tau$ and longitudinal
coordinate $\zeta$, which are related with time $t$ and space $z$
through $\tau =\sqrt{\gamma_{||}\gamma_{\bot}}t$ and $\zeta=2\pi
z/\left(  \tilde{\alpha }L_{\mathrm{m}}\right)  $, where
\begin{equation}
\tilde{\alpha}=\frac{2\pi
c}{L_{\mathrm{c}}\sqrt{\gamma_{\Vert}\gamma_{\bot}}}
\label{alfa}%
\end{equation}
is the adimensional free spectral range of the cavity, being $c$
the light velocity in the host medium. The actual free spectral
range of the cavity, FSR, is related to $\tilde\alpha$ by
\begin{equation}
\mathrm{FSR}=\frac{2\pi c}{L_{\mathrm{c}}}=\left(
\gamma\tilde{\alpha
}\right)  \gamma_{\bot}, \label{FSR}%
\end{equation}
hence $\gamma\tilde{\alpha}$ represents the FSR measured in units
of the homogeneous linewidth. The boundary condition for the
electric field
\begin{equation}
F(0,\tau)=F\left(  {2\pi}/{\tilde{\alpha}},\tau\right)
\end{equation}
means that $F$ can be expressed as a superposition of plane waves
with a spatial wave--number $\alpha$ equal to an integer multiple
of $\tilde{\alpha}$ (with our scaling of space and time $\alpha$
denotes both the spatial wave--number and the temporal frequency).

Eqs. (\ref{mod1}-\ref{mod3}) have two stationary solutions, the
laser--off solution $\bar{F}=\bar{P}=0$ and $\bar{D}=1$, and the
resonant single--mode lasing solution
$\bar{F}=\sqrt{A-1}\,\mathrm{e}^{i\phi}$, $\bar{P}=\bar{F}/A$ and
$\bar{D}=1/A$, where $\phi$ is an arbitrary phase. This solution
appears at the lasing threshold $A=1$.

The linear stability analysis of the single--mode solution has
been reported many times, see e.g., \cite{Roldan05}. This solution
is unstable for a given $A$ if $\alpha_{-}<\alpha<\alpha_{+}$ with
\begin{eqnarray}
\alpha_{\pm}  &=& \omega_{\pm}\left(  1+\frac{\gamma\sigma}{A-\omega_{\pm}%
^{2}}\right)  \,,\label{lsa1}\\
\omega_{\pm}^{2}  &=& \frac{1}{2}\left[
3(A-1)-\gamma^{2}\pm\sqrt{R}\right]
\,,\label{lsa2}\\
R  &=& (A-1)(A-9)-6\gamma^{2}(A-1)+\gamma^{4}\,. \label{lsa3}%
\end{eqnarray}
These equations give the bifurcation line on the $\left\langle
A,\alpha \right\rangle $ plane. The instability occurs for a
minimum pump when $A=A_{c}$ and $\alpha=\alpha_{c}$ where
\[
A_{c}=5+3\gamma^{2}+2\sqrt{2}\sqrt{\left(  \gamma^{2}+2\right)
\left( \gamma^{2}+1\right)  }\,,
\]
and $\alpha_{c}$ can be obtained from Eqs. (\ref{lsa1}) and
(\ref{lsa2}) setting $R=0$ and $A=A_{c}$. In Fig. \ref{fig:insta},
we represent the instability boundary on the $\left\langle
A,\alpha\right\rangle $ plane for $\gamma=1$ and $\sigma=0.05$.

\section{Numerical results}

We have numerically integrated Eqs. (\ref{mod1}-\ref{mod3}) for
fixed relaxation rates $\gamma=1$ and $\sigma=0.05$, letting the
frequency $\alpha$ and the pump $A$ as variable parameters. Notice
that, although $\sigma \ll\gamma$, this choice of the parameters
does not mean that we are considering a class--A laser,
\textit{i.e.} a laser where the time evolution of electric field
is much slower than those of the material variables. This would be
true in the single--mode limit, but here we are studying a
multi--mode laser, where the $n$--th side--mode oscillates at an
angular frequency $\alpha_{n}=n\,\tilde{\alpha}$ of order 1 or
larger, therefore the time scale of the electric field is
comparable to those of the material variables, as in class--C
lasers.

Our choice of $\sigma$ corresponds to a mirror transmissivity $T$
close to 0.1. In fact, if distributed losses are negligible, the
cavity linewidth $\sigma$ and the free spectral range
$\tilde{\alpha}$ are related by $\sigma=\tilde{\alpha}\,T/(2\pi)$,
and we will consider values of $\tilde{\alpha}$ around 4.

The integration method is based on a modal expansion of the
electric field \cite{deValcarcel03a}
\begin{equation}
F(\zeta,\tau)=\sum_{n=-N}^{N}\mathrm{e}^{i\,n\tilde{\alpha}\zeta}f_{n}%
(\tau)\,,
\end{equation}
which allows to convert Eqs. (\ref{mod1}-\ref{mod3}) into a set of
ordinary integro--differential equations for the $2N+1$ complex
mode amplitudes $f_{n}$ and for the variables
$P_{m}(\tau)=P(\zeta_{m},\tau)$ and $D_{m}(\tau
)=D(\zeta_{m},\tau)$, with $m=1\ldots M$. For the present analysis
we verified that 11 modes ($N=5$) and a spatial grid of $M=21$
points are enough to reproduce accurately the total electric
field.

\begin{figure}[ptb]
\centering {\scalebox{.7}{\includegraphics*{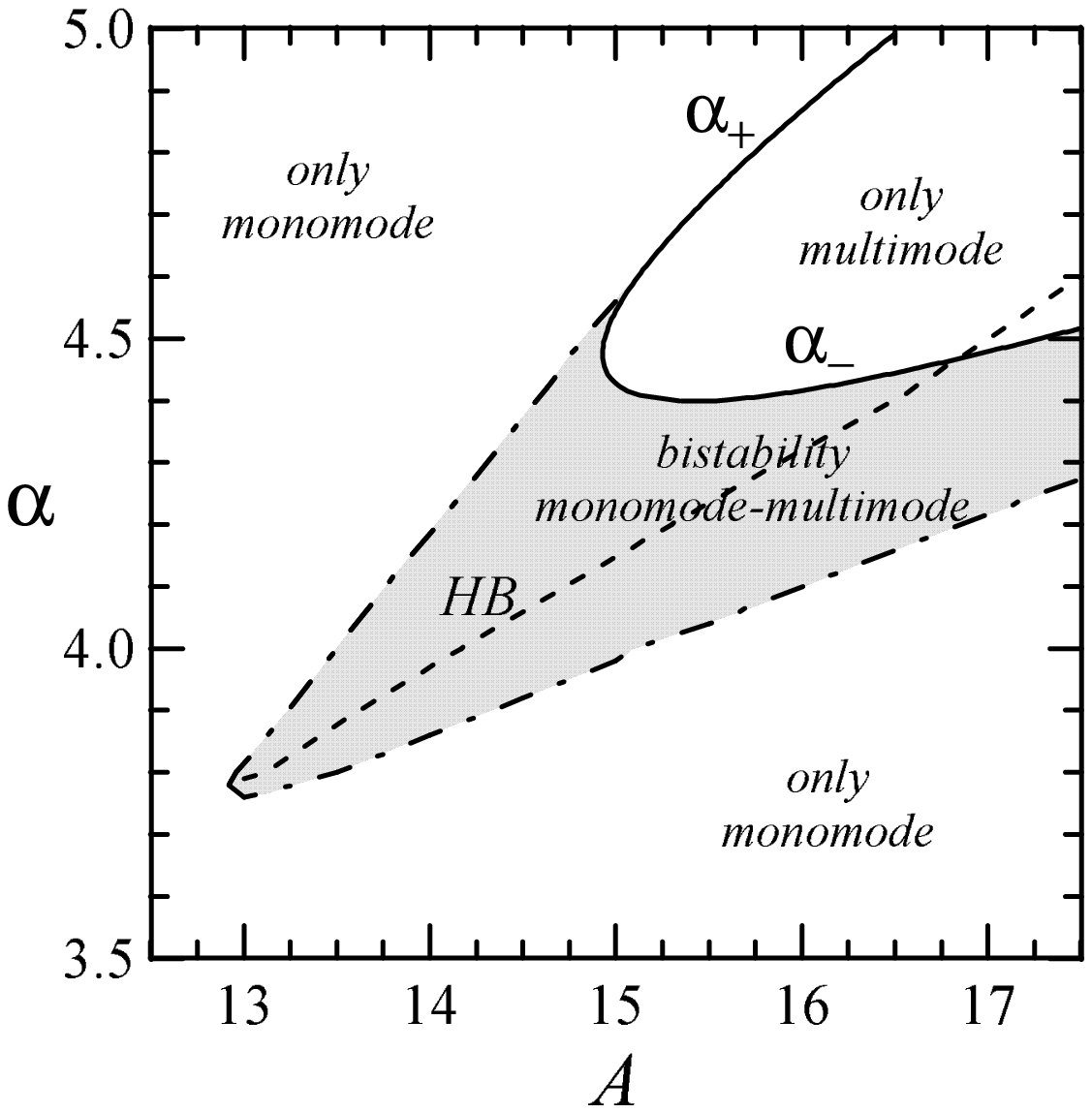}}}\caption{The
instability domain of Fig. \ref{fig:insta} is represented together
with the existence domain of the multi--mode solution, delimited
by the dashed--dotted line. In the shaded area there is
bistability between the single-- and the multi--mode solution. The
dashed line indicates the threshold for the Hopf bifurcation of
the multi--mode solution.} \label{fig:bista}\vglue1cm
\end{figure}

We proceeded as follows: First we fixed $\alpha$ between
$\alpha_{-}$ and $\alpha_{+}$ and took a pump value for which the
single--mode solution is unstable. Then we varied the pump $A$ in
both directions to determine the boundaries of the multi--mode
solution. We repeated the operation for several values of $\alpha$
even moving below $\alpha_{-}$. In Fig. \ref{fig:bista} the
boundary of multi--mode emission found in this way is represented
with a dashed--dotted line together with the boundary of the
single--mode solution instability domain, indicated by the solid
line. The shadowed area marks the domain where both the
single--mode and the multi--mode solutions are stable.

Two features of the bistability domain are of particular
interest:\newline(i)
The domain extends to the left up to $A\simeq13$, well below $A_{c}%
=14.93$.\newline(ii) The domain extends up to a frequency $\bar{\alpha}%
\simeq4.56$ larger than $\alpha_{c}=4.47$. Hence, considering $A$
as the bifurcation parameter, the bifurcation is subcritical not
only for $\alpha_{\mathrm{min}}<\alpha<\alpha_{c}$, as shown by Fu
in the class--B limit, but also for
$\alpha_{c}<\alpha<\bar{\alpha}$.\newline(iii) The domain extends
up to a minimum frequency $\alpha_{\min}\approx3.76$.

Let us now characterize the multi--mode dynamics existing in the
range of parameters covered in Fig. \ref{fig:bista}. There are two
clearly different regimes that are separated by the dashed line
that crosses the bistability regime (marked as $HB$). At the left
of the dashed line, multi--mode emission is periodic, the modal
intensities are constant and phases are locked. The dashed line
marks a Hopf bifurcation. At the right of this line the dynamics
of the total intensity is quasi--periodic and the modal
intensities and the relative phases oscillate periodically in
time.

\begin{figure}[ptb]
\centering {\scalebox{.7}{\includegraphics*{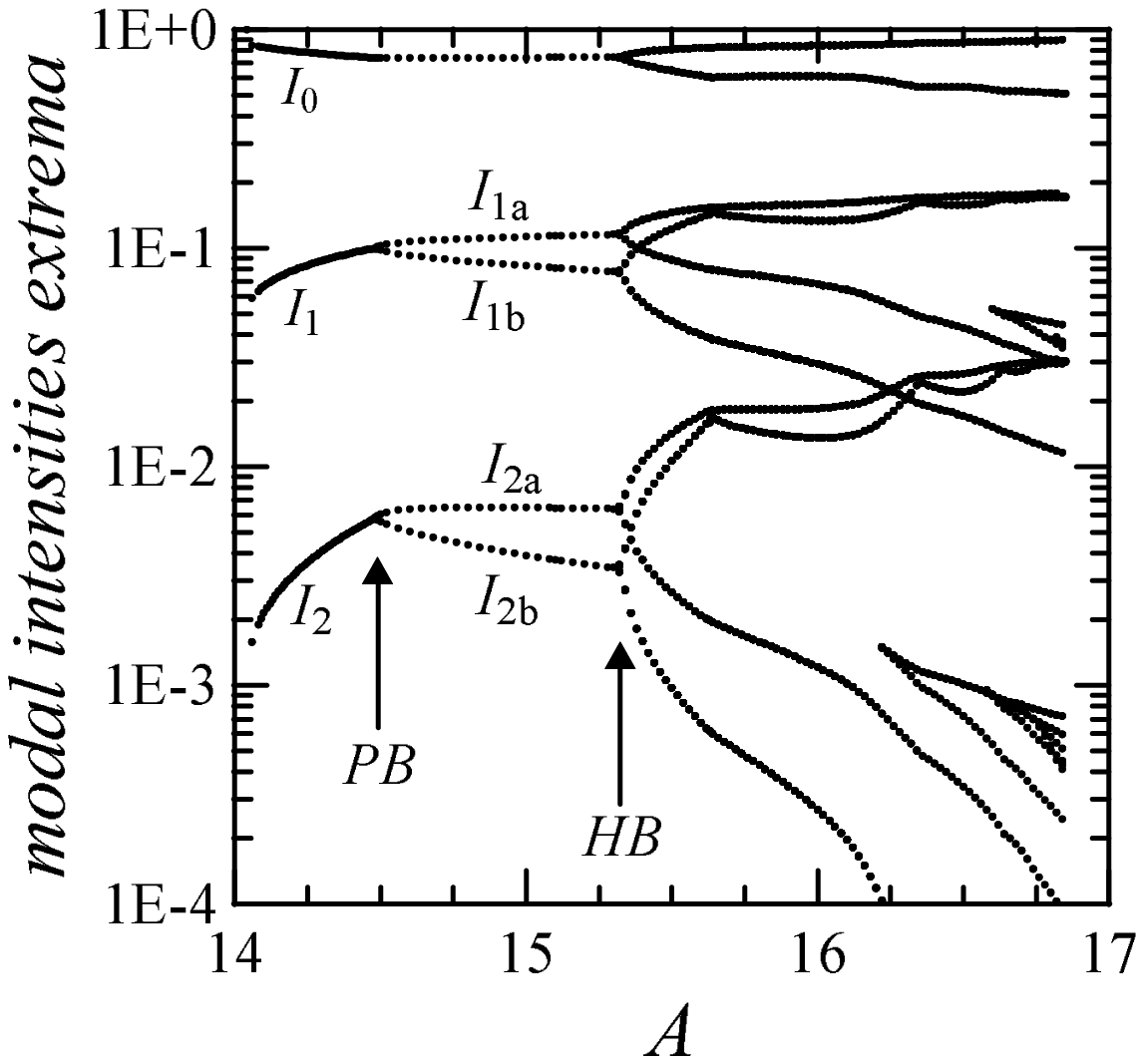}}}\caption{Modal
intensities as a function of the pump $A$ for the central modes
and the first two side--modes. The parameters are $\sigma=0.05$,
$\gamma=1$, and $\alpha=4.2$. Beyond the Hopf bifurcation the
intensities are no longer constant, and the lines indicate the
extrema of the oscillations.} \label{fig:int-A}\vglue1cm
\end{figure}

In order to analyze this in more detail, we describe now the
dynamics of the system for the particular value $\alpha=4.2$ as a
function of the pump parameter $A$. In Fig. \ref{fig:int-A} the
modal intensities corresponding to the central mode, $I_{0}$, and
the two first sidebands (modes $I_{\pm1}$ and $I_{\pm2}$) are
represented as a function of $A$ (the intensities of higher order
modes are, at least, one order of magnitude smaller than
$I_{\pm2}$).

There are three clearly distinguishable zones in Fig.
\ref{fig:int-A}:\newline(i) For $A<A_{PB}=14.49$, the modal
intensities are constant and single--valued, and symmetric modes
have the same intensity ($I_{+n}=I_{-n}$,
$n=1\ldots5$).\newline(ii) for $A_{PB}<A<A_{HB}=15.33$, the modal
intensities are constant but there are two possible solutions,
denoted as $I_{na}$ and $I_{nb}$ in the figure, and symmetric
modes have different intensities ($I_{+n}\neq I_{-n}$). Precisely,
when $I_{+n}=I_{na}$, then $I_{-n}=I_{nb}$, and
viceversa.\newline(iii) For $A>A_{HB}$ the modal intensities are
no more constant (for this domain, in the figure we have
represented the extrema of the modal intensities).

Hence, looking at the modal intensities, we can conclude that they
are subject to a pitchfork, symmetry breaking, bifurcation at
$A=A_{PB}$ and to a Hopf bifurcation at $A=A_{HB}$.

We studied more in detail the two multi--mode self--pulsing
solutions that coexist in the interval $A_{PB}<A<A_{HB}$. One
could imagine that these solutions differ in the total intensity,
because all the modes placed on one side of the spectrum have
intensities larger than the corresponding modes on the other side.
But this is not the case: the shape of the pulses emitted by the
laser is exactly the same for the two solutions. How this can be
possible can be understood looking at the behavior of modal
frequencies and phases.

\begin{figure}[ptb]
\centering {\scalebox{.8}{\includegraphics*{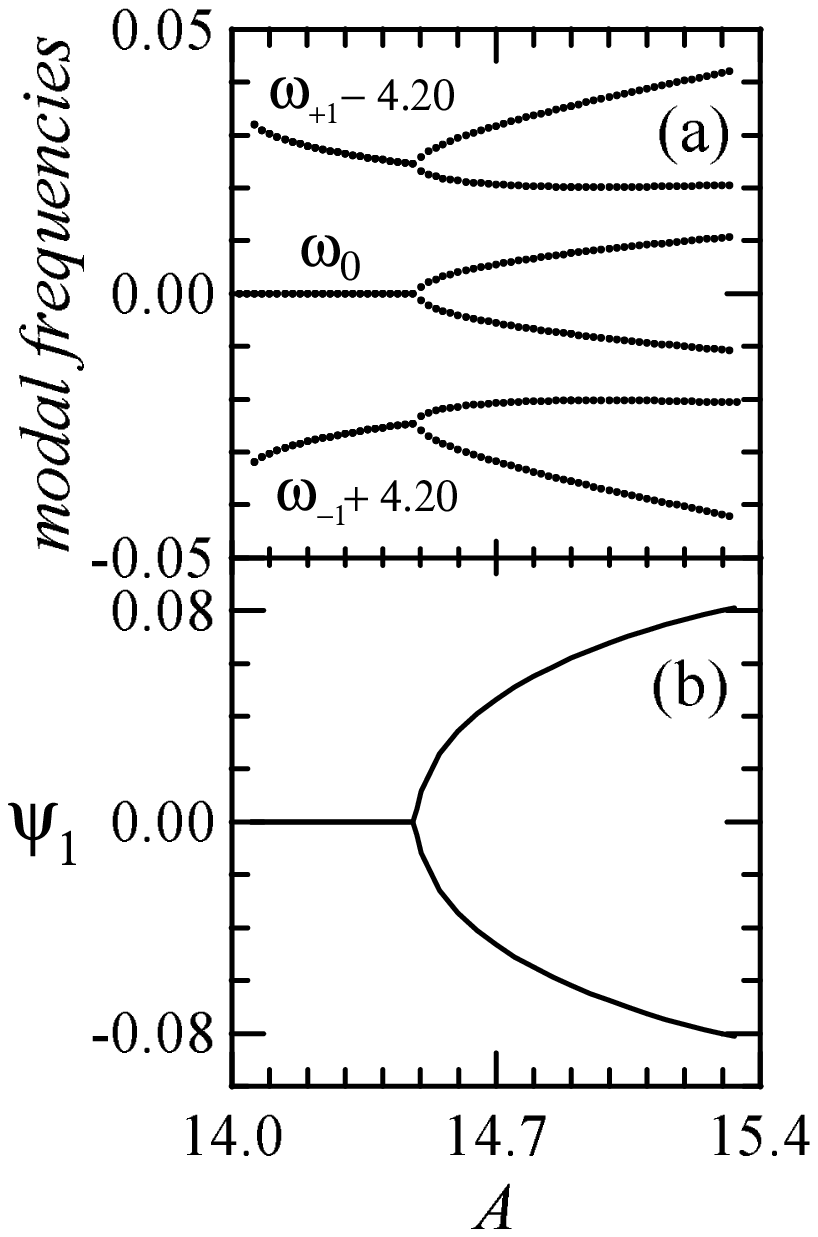}}}\caption{Modal
frequencies (a) and relative phase (b) as a function of the pump
$A$ in the region around the pitchfork bifurcation. The relative
phase $\Psi_{1}$ is measured in radians.}
\label{fig:frefas}\vglue1cm
\end{figure}

In the upper panel of Fig. \ref{fig:frefas} we show the calculated
frequencies of the central mode $(\omega_{0})$ and of the first
side--modes $(\omega_{\pm1})$, to which we have subtracted the
empty cavity frequencies $\pm\alpha$. In the lower panel of the
same figure we represent in radians the relative phase
$\Psi_{1}=\phi_{+1}+\phi_{-1}-2\phi_{0}$, where $\phi_{i}$ is the
phase of the $i$--th mode. If the solution is phase--locked
$\Psi_{1}$, as well as all the other relative phases $\Psi_{i}$
that can be defined in the same way, must be constant.
\begin{figure}[ptb]
\centering {\scalebox{.8}{\includegraphics*{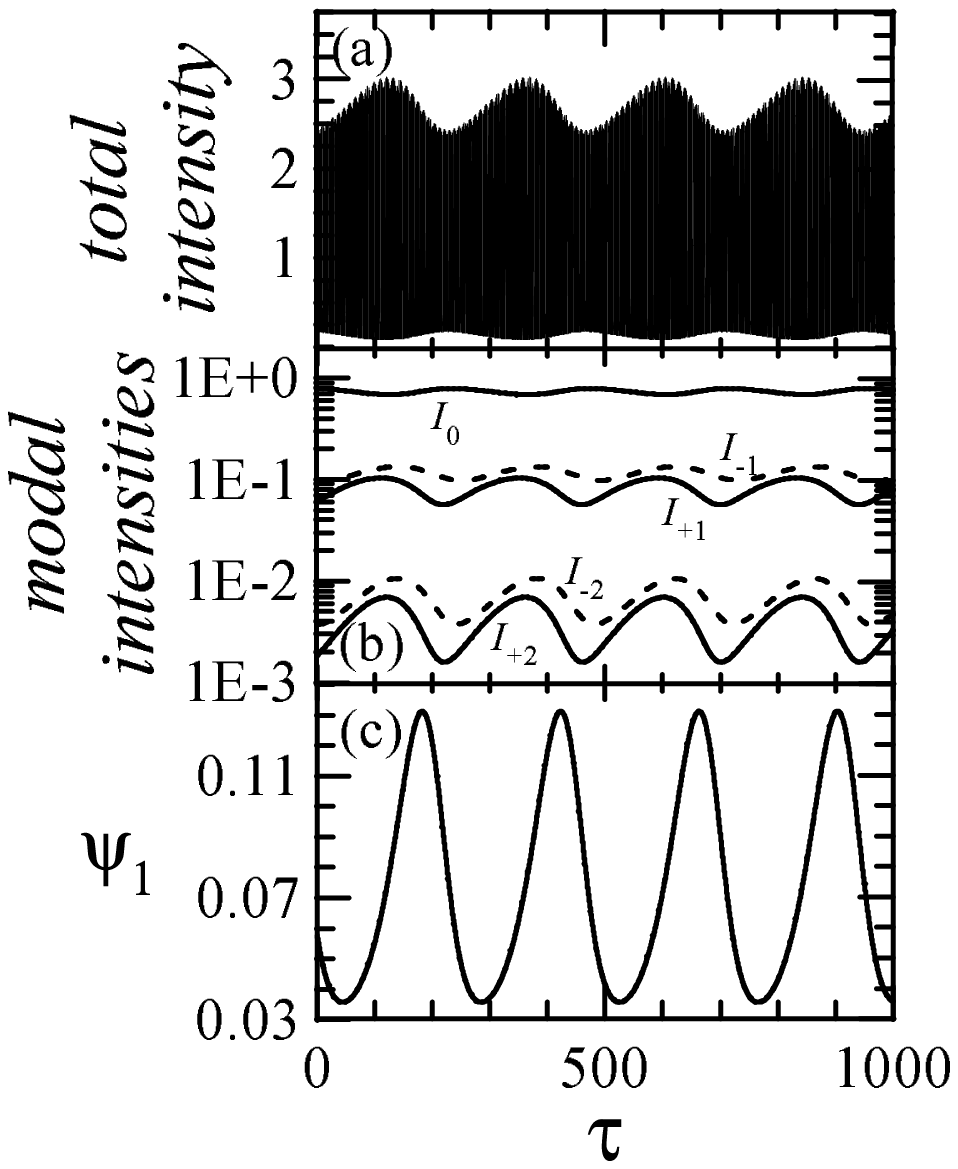}}}\caption{Total
intensity (a), intensities of the central mode and of the first
two side--modes (b), and relative phase $\Psi_{1}$ (c) after the
Hopf bifurcation. $A=15.4$} \label{fig:15.4}\vglue1cm
\end{figure}
\begin{figure}[ptb]
\centering {\scalebox{.8}{\includegraphics*{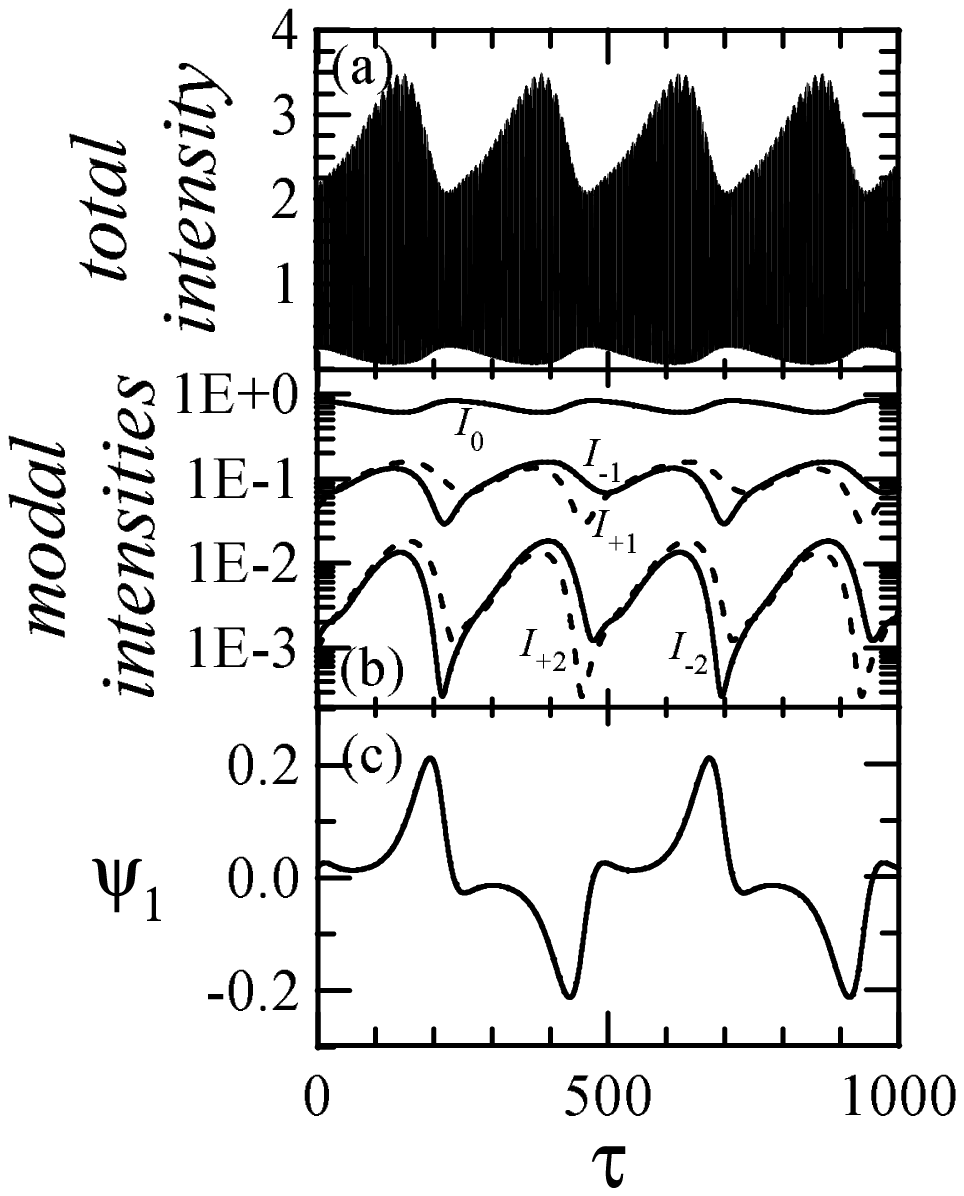}}}\caption{Same
as Fig. \ref{fig:15.4} for $A=16.0$.} \label{fig:16.0}\vglue1cm
\end{figure}
\begin{figure}[ptb]
\centering {\scalebox{.8}{\includegraphics*{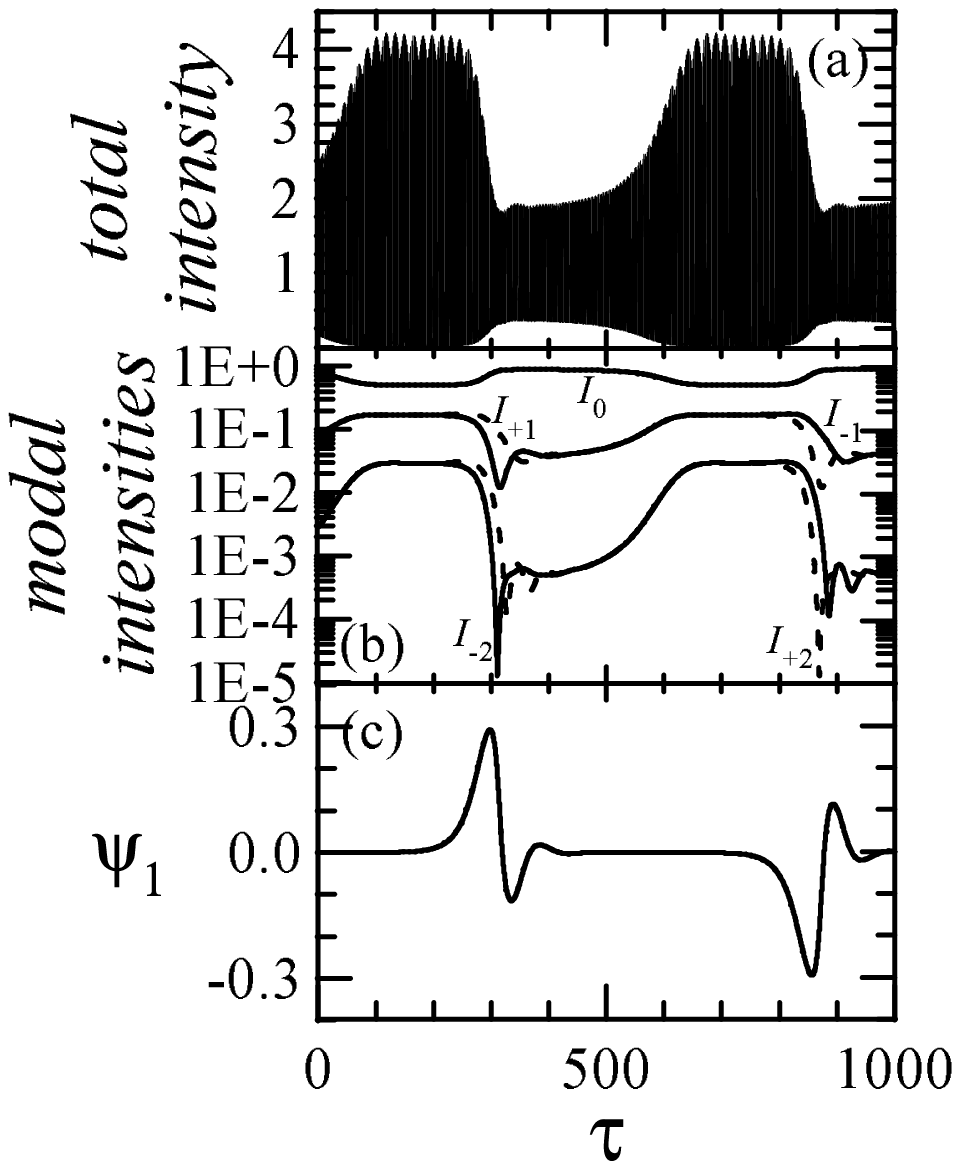}}}\caption{Same
as Fig. \ref{fig:15.4} for $A=16.8$.} \label{fig:16.8}\vglue1cm
\end{figure}

We see that at the pitchfork bifurcation the modal frequencies
experience a shift in the same direction, positive or negative for
the two solutions. This shift however preserves phase--locking,
although $\Psi_{1}$ is no longer equal to 0 as for $A<A_{PB}$, but
it can take two opposite values, associated with the two
solutions. Notice that the maximum frequency separation between
the two phase--locked solution is about 0.02.

The combination of different mode intensities, frequencies and
relative phases makes it possible that the total intensities for
the two solutions are identical.

Let us now analyze the laser dynamics beyond the Hopf bifurcation
$A=A_{HB}$. In Figs. \ref{fig:15.4}, \ref{fig:16.0}, and
\ref{fig:16.8} we considered the three different values for the
pump $A=15.4$, $A=16$ and $A=16.8$. In each figure the upper panel
shows the total intensity, the middle panel the modal intensities
of the central mode and of the first two side--modes, and the
lower panel the relative phase $\Psi_{1}$ defined above, measured
in radians. The total intensity displays a slow modulation with
period of some hundreds time units, superimposed to the much
faster self--pulsing oscillations of period
$2\pi/\alpha\approx1.5$. This slow modulation is clearly related
to the oscillations of the modal intensities and of the relative
phase, which have the same period. The period is almost the same
in Figs. \ref{fig:15.4} and \ref{fig:16.0}, and it is almost twice
larger in Fig. \ref{fig:16.8}. The strength of the modulation
increases with $A$, and the relative phase $\Psi_{1}$ passes from
the almost regularly periodic oscillations of Fig.
\ref{fig:15.4}(c) to the behavior shown in Fig. \ref{fig:16.8}(c),
where $\Psi_{1}$ remains most of the time close to 0.

We notice that the slow modulation frequency for the smaller
values of $A$ is close to the maximum frequency separation between
the two phase--locked solutions that is achieved immediately
before $A=A_{HB}$. Hence, we may interpret the dynamical state
that arises from the Hopf bifurcation as a state where the two
phase--locked solutions are present simultaneously, and oscillate
at their beat note.

Let us finally comment that we have not found chaotic behavior.
Certainly the quasi--periodic dynamics of the total intensity is
more involved as pump increases, but multi--mode emission
disappears before more complex dynamics develops.

\section{Conclusion}

We have numerically and theoretically investigated bistability
between single--mode and multi--longitudinal mode solutions in the
standard ring--cavity two--level laser within the uniform field
limit. We have determined the domain of coexistence between the
single--mode and multi--longitudinal mode solutions for a class--C
laser (we have used $\gamma=1$, and $\sigma=0.05$) finding that
this domain is relatively wide. In particular we have found that
the domain of coexistence is different from that corresponding to
a class--B laser as it extends for $\alpha$ slightly larger than
$\alpha_{c}$ and for $A<A_{c}$ (for class--B lasers it exists for
$\alpha\leq\alpha_{c}$ and $A\geq A_{c}$ \cite{Fu89,Carr94a}). We
have also found that the multimode solution undergoes a pitchfork
bifurcation (which is a symmetry breaking bifurcation) and a
subsequent Hopf bifurcation that destroys mode--locking. In the
near future we plan to extend this numerical study to a situation
closer to that of the experimental conditions in \cite{Voigt04},
in order to determine up to what extent the deviations from the
theoretial predictions could be interpreted as a manifestation of
the coexistence between single--mode and multi--longitudinal mode
emission.

We gratefully acknowledge G.J. de Valc\'{a}rcel for continued
discussions. This work has been supported by the Spanish
Ministerio de Ciencia y Tecnolog\'{\i}a and European Union FEDER
(Fonds Europ\'{e}en de D\'{e}velopppement R\'{e}gional) through
Project PB2002-04369-C04-01.



\end{document}